\documentclass[amsmath, amssymb, aps, pra, reprint, twocolumn, longbibliography, floatfix, nofootinbib, superscriptaddress]{revtex4-2}
\usepackage[utf8]{inputenc}
\usepackage[urlcolor=blue, colorlinks=true, citecolor=blue]{hyperref}
\usepackage[english]{babel}
\usepackage{amsmath}
\usepackage{amsmath, amssymb}
\usepackage{nicematrix}
\usepackage{textcomp, gensymb}
\usepackage{tabularx}
\usepackage{subcaption}
\usepackage{algorithm}
\usepackage{physics}
\usepackage{algpseudocode}
\usepackage{comment}
\usepackage{caption}
\usepackage[euler]{textgreek}

\begin{document}
\title{Numerical solution of the incompressible Navier-Stokes equations for chemical mixers via quantum-inspired Tensor Train Finite Element Method}

\author{Egor~Kornev}
\address{Terra Quantum AG, Kornhausstrasse 25, 9000 St. Gallen, Switzerland}

\author{Sergey~Dolgov}
\address{Terra Quantum AG, Kornhausstrasse 25, 9000 St. Gallen, Switzerland}


\author{Karan~Pinto}
\address{Terra Quantum AG, Kornhausstrasse 25, 9000 St. Gallen, Switzerland}

\author{Markus~Pflitsch}
\address{Terra Quantum AG, Kornhausstrasse 25, 9000 St. Gallen, Switzerland}

\author{Michael~Perelshtein}
\email{mpe@terraquantum.swiss}
\address{Terra Quantum AG, Kornhausstrasse 25, 9000 St. Gallen, Switzerland}

\author{Artem~Melnikov}
\address{Terra Quantum AG, Kornhausstrasse 25, 9000 St. Gallen, Switzerland}

\begin{abstract}
The solution of computational fluid dynamics problems is one of the most computationally hard tasks, especially in the case of complex geometries and turbulent flow regimes. 
We propose to use Tensor Train (TT) methods, which possess logarithmic complexity in problem size and have great similarities with quantum algorithms in the structure of data representation.
We develop the {\tt Te}nsor {\tt tra}in {\tt F}inite {\tt E}lement {\tt M}ethod -- {\tt TetraFEM} -- and the explicit numerical scheme for the solution of the incompressible Navier-Stokes equation via Tensor Trains. 
We test this approach on the simulation of liquids mixing in a T-shape mixer, which, to our knowledge, was done for the first time using tensor methods in such non-trivial geometries. 
As expected, we achieve exponential compression in memory of all FEM matrices and demonstrate an exponential speed-up compared to the conventional FEM implementation on dense meshes. 
In addition, we discuss the possibility of extending this method to a quantum computer to solve more complex problems.
This paper is based on work we conducted for Evonik Industries AG.

\end{abstract}

\maketitle

\section{INTRODUCTION}

The modelling of fluid dynamics is a demanding and challenging field \cite{en15228405, Claudio_Chiastra, HAN2022104485}. 
The partial differential equations (PDEs) describing the physical processes are complex and in an overwhelming number of practical cases, are not analytically solvable.
This means numerical approaches are required \cite{navier-survey}. 
The main limitation of these approaches is the inevitable trade-off between computational resources 
and the quality of the solution. 
Producing satisfactory results usually requires very dense discretization, which leads to enormous amounts of memory and processor runtime being consumed by the algorithm, especially for transient flows \cite{navier-num-survey}.
To overcome these challenges, alternative methods for approximating flow functions with lower computational complexity are highly sought.

One such promising and novel approach is based on Tensor Networks. 
This mathematical tool was first introduced in the context of multi-particle quantum physics \cite{orus2019tensor_networks}. 
Independently, it was proposed for the efficient solution of computational mathematical problems: basic linear algebra \cite{TT_main_paper}, including solution of linear systems of equations \cite{AMEN, Multigrid}, optimization \cite{sozykin2022ttopt, morozov2023protein}, and machine learning \cite{novikov2015tensorizing, sagingalieva2022hyperparameter, naumov2023tetra-aml}. 
The main advantage of tensor network algorithms is an exponential reduction in storage and a poly-logarithmic runtime complexity by the mesh discretization size compared to conventional methods \cite{blazek2015cfd}.
For example, this has been demonstrated on a number of PDE problems \cite{dolgov2012fast,dp-chemotaxis-2019, Variational, gourianov2022quantum, Thesis}.
In addition, an important feature of tensor networks is that they can be quite efficiently mapped to a quantum computer \cite{MPS_preparation}.

Unfortunately, all of the problems mentioned above were solved in simple rectangle-type domains in order for the function values on the mesh to be well-represented as tensors.
The purpose of this work is to apply the tensor network approach to more complex T-mixer-type domains, similarly to Refs.~\cite{bunger2020_isogeometric, markeeva2021qtt}.

Here, we want to emphasize why we don't use quantum PDE solvers despite the large variety of them \cite{gaitan2020_Nav_St_quantum, Poisson_on_NISQ, childs2021_pde}.
Quantum algorithms usually solve a PDE problem by reducing it to the solution of a linear system of equations and applying the famous HHL \cite{harrow2009quantum, perelshtein2022solving} or more advanced algorithms, e.g. Child's \cite{childs2017_linear_systems}. 
Although these algorithms can theoretically be applied to solving PDEs, in practice there are many challenges -- the main one is the great complexity and inefficiency of encoding unitary matrices into a quantum computer \cite{krol2022_unitary_decomposition}. 
Therefore, basically, very small tasks are solved with the help of these algorithms, and, as Ref.~\cite{Rolls_roys_CFD} shows, these methods are very poorly scalable. 
In this regard, we adhere to a different approach -- we first try to solve the problem using algorithms on tensor networks, which can be realized on a classical computer, and then try to extend them to a quantum computer. 
In addition to the immediate complexity reduction on a classical computer in comparison to conventional methods, this approach allows us to better understand which problems can benefit from hybrid quantum computing \cite{white_paper_tq} and how.

The structure of the paper is as follows.
In Section~\ref{TT_algorithms}, we give an overview of Tensor Trains and the main algorithms.
The main ideas behind the finite element analysis in application to the problem are presented in Section~\ref{TetraFEM}. 
In Section~\ref{equations}, the model equations and the mathematical problem statement are presented. 
Section~\ref{test} provides the results of the numerical tests.
In Section~\ref{quantum_extension}, we discuss a possible extension of the presented method to a quantum computer and Section~\ref{conclusion} concludes the work.

\section{Tensor Trains}\label{TT_algorithms}

Tensor Trains (TTs) are the basic type of Tensor Network on which all basic linear algebra operations can be efficiently realized \cite{TT_main_paper}. 
In this section, we will give a definition of Tensor Trains and describe the main TT-algorithms used in our implementation, which are {\tt Rounding} and {\tt MatVec} operations \cite{TT_main_paper}, as well as solving a linear system of equations in TT format \cite{AMEN} and TT-cross approximation \cite{tt_cross}. In addition, we consider the Quantized Tensor Train format in its canonical form \cite{QTT}.
Table \ref{table:operations} contains the most important information about the operations that are used.

\begin{figure}[h!]
        \includegraphics[width = 1 \linewidth]{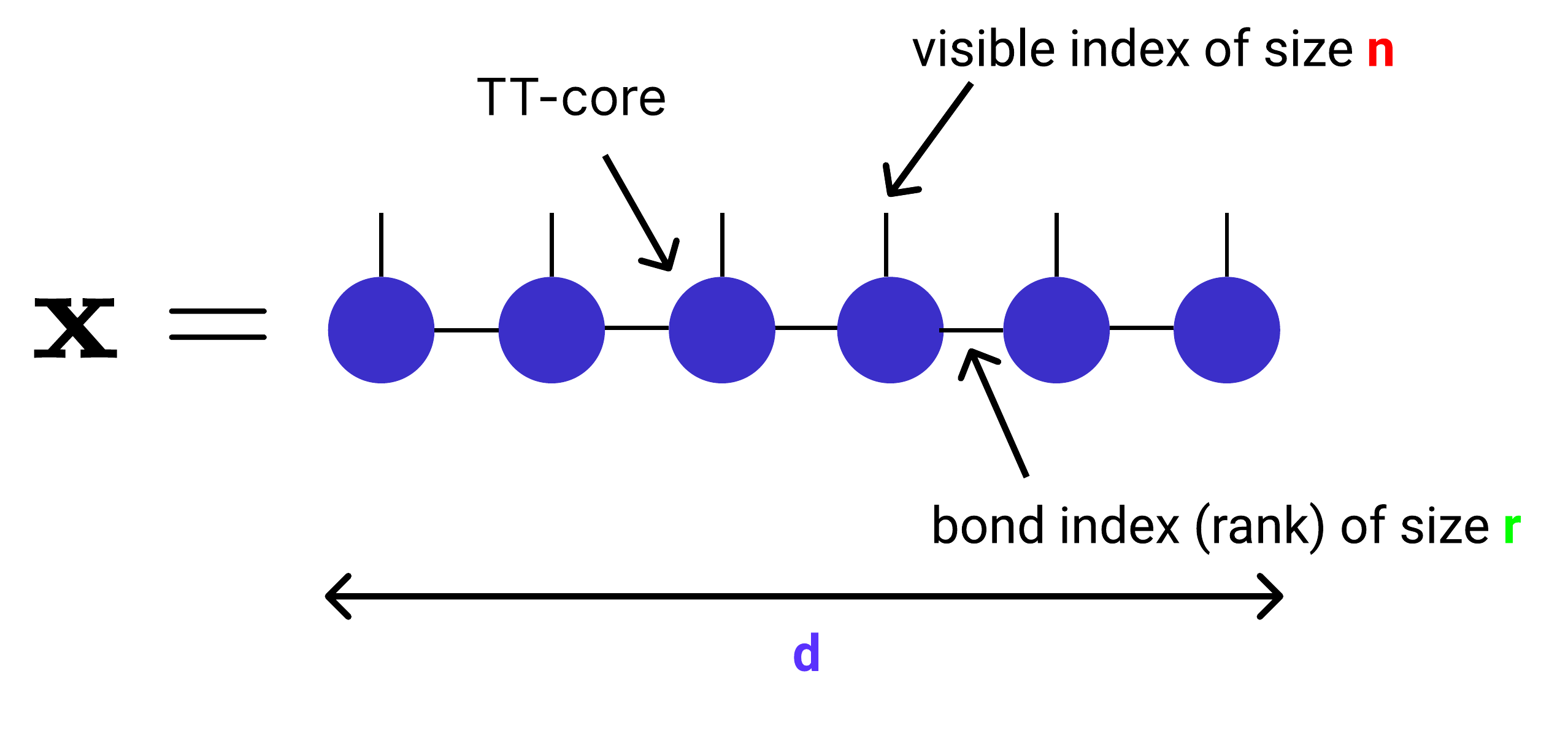}
        \caption{Representation of a $d$-dimensional $n\times \cdots \times n$ tensor of rank $r$ in a TT format. 
        Shapes with ``legs'' (TT-cores) represent tensors of dimension equal to the number of legs. 
        The connection of the tensors along the leg denotes their convolution by this index.}
        \label{MPS}
\end{figure}

\begin{table}[h!]
\[\begin{array}{clll}
\hline 
\text { № } &\text { Operation } & \text { Result rank  } & \text { Complexity } \\
\hline 
1 & \mathbf{z}=\mathbf{x} \cdot \text { const } & \mathrm{r}(\mathbf{z})=\mathrm{r}(\mathbf{x}) & O(d \mathrm{r}(\mathbf{x})) \\

2 &\mathbf{z}=\mathbf{x}+\mathbf{y} & \mathrm{r}(\mathbf{z}) \leq \mathrm{r}(\mathbf{x})+\mathrm{r}(\mathbf{y}) & O\left(n d(\mathrm{r}(\mathbf{x})+\mathrm{r}(\mathbf{y}))^{2}\right) \\

3 & \mathbf{z}=\mathbf{x} \odot \mathbf{y} & \mathrm{r}(\mathbf{z}) \leq \mathrm{r}(\mathbf{x}) \mathrm{r}(\mathbf{y}) & O\left(n d \mathrm{r}^{3}(\mathbf{x}) \mathrm{r}^{3}(\mathbf{y})\right) \\

4 & \operatorname{matvec} \mathbf{z}= \mathcal{A} \mathbf{x} & \mathrm{r}(\mathbf{z}) \leq \mathrm{r}(\mathcal{A}) \mathrm{r}(\mathbf{x}) &  O\left(n d \mathrm{r}^{3}(\mathcal{A}) \mathrm{r}^{3}(\mathbf{x})\right) \\

5 &\operatorname{solve} \mathcal{A} \mathbf{x} =  \mathbf{y} & r(\mathbf{x}) & O\left(n d \mathrm{r}^{3}(\mathbf{x}) \mathrm{r}(\mathcal{A})\right) \\

6 &\mathbf{z}=\operatorname{round}(\mathbf{x}, \varepsilon) & \mathrm{r}(\mathbf{z}) \leq \mathrm{r}(\mathbf{x}) & O\left(n d \mathrm{r}^{3}(\mathbf{x})\right) \\
\hline
\end{array}\]
\caption{$\mathbf{x}$, $\mathbf{y}$ and $\mathbf{z}$ are tensor train vectors of the same dimensions. Their ranks are $\mathrm{r}(\mathbf{x})$,  $\mathrm{r}(\mathbf{y})$ and $\mathrm{r}(\mathbf{z})$, respectively. $\mathcal{A}$ is a tensor train matrix of rank $\mathrm{r}(\mathcal{A})$. (1) - multiplication of a vector by a constant, (2) - element-wise addition of two vectors, (3) - element-wise product of two vectors, (4) - matrix-vector multiplication, (5) - solution of a linear system of equations, (6) -  rounding that effectively reduces the rank of a tensor train with a given precision.}
\label{table:operations}
\end{table}

\subsection{Definition}
A Tensor Train can be considered an effective representation of multidimensional arrays \cite{TT_main_paper}. 
Such a decomposition is given by
\begin{align*}
    \mathbf{x}\left(i_{1}, \ldots, i_{d}\right)= \sum_{\alpha_{0}, \ldots, \alpha_{d-1}, \alpha_{d}} & G_{1}\left(\alpha_{0}, i_{1}, \alpha_{1}\right) G_{2}\left(\alpha_{1}, i_{2}, \alpha_{2}\right) \\ & \ldots G_{d}\left(\alpha_{d-1}, i_{d}, \alpha_{d}\right),
\end{align*}
where $G_j$ is the 3-dimensional tensor called the TT-core.
The main characteristic of such a representation is the rank, $r$, which is equal to the maximum size among the indices $\alpha_{0}, \alpha_{1}, \ldots, , \alpha_{d}$ and expresses the correlations and the entanglement in the tensor.
In the case of weak correlations (low rank $r$), this format allows one to store and perform operations with tensors with logarithmic complexity in their size.
The graphical representation of a tensor train is presented in Fig.~\ref{MPS}.

\begin{figure}[h!]
        \includegraphics[width = 1 \linewidth]{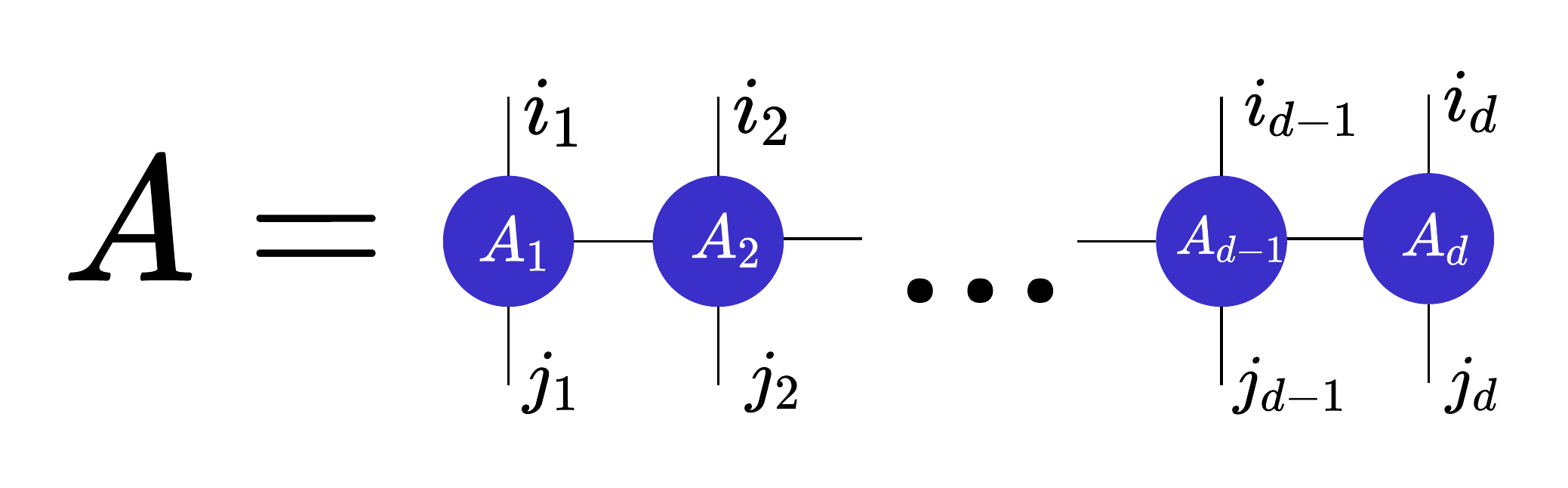}
         \caption{Graphical representation of a TT-Matrix $\mathcal{A}$. 
         The blue shapes with legs denote TT-cores.
         $\left(i_1, \ldots, i_d\right)$ enumerates the rows of $\mathcal{A}$ and $\left(j_1, \ldots, j_d\right)$ enumerates its columns.}
         \label{MPO}
\end{figure}

We define a \emph{TT-vector} as any tensor in the form $\mathbf{x}\left(i_{1}, \ldots, i_{d}\right)$.
We implicitly assume that a vector of length $N = n_1 \ldots n_d$ is treated as a $d$-dimensional tensor with mode sizes $n_k$.
This tensor is represented in the TT format as shown in Fig.~\ref{MPS}.
Following similar reasoning, we define a \emph{TT-matrix} as any tensor of the form $\mathcal{A}\left(i_1, \ldots, i_d, j_1, \ldots, j_d\right)$, where $\left(i_1, \ldots, i_d\right)$ enumerates the rows of $\mathcal{A}$ and $\left(j_1, \ldots, j_d\right)$ enumerates its columns. 
A graphical representation of a TT-matrix is presented in Fig.~\ref{MPO} and it reads as $\mathcal{A} = A_1\left(i_1, j_1\right) \ldots A_d\left(i_d, j_d\right)$.

\subsection{Algorithms}
Basic linear algebra operations, such as multiplication of a TT-vector by a constant, addition of two TT-vectors, and element-wise multiplication, can be effectively implemented in TT format (the information is provided in the Table~\ref{table:operations}) \cite{TT_main_paper}. 
As one can see, after applying these operations, the rank of the result can increase significantly, so it needs to be effectively lowered after each step - the {\tt Rounding} operation serves this purpose.
The {\tt Rounding} operation allows one to efficiently reduce the rank of a given TT-vector $\mathbf{x}$ with a specified accuracy. 
The complexity of this operation is $O(dnr(\mathbf{x})^3)$, where $r(\mathbf{x})$ is the initial TT-rank of $\mathbf{x}$.

The {\tt MatVec} operation is a multiplication of a TT-vector, $\mathbf{x}$, by a TT-matrix, $\mathcal{A}$, which gives the TT-vector, $\mathbf{y} = \mathcal{A} \mathbf{x}$ \cite{TT_main_paper}.
The straightforward implementation, where one collects TT-cores of $\mathbf{y}$ just as a product of $\mathbf{x}$ and $\mathcal{A}$ TT-cores and then performs {\tt Rounding} to decrease the ranks has complexity $O(dnr(\mathbf{x})^3r(\mathcal{A})^3)$. 

Another important operation we used is the solution of a linear system of equations $\mathcal{A} \mathbf{x}= \mathbf{b}$ via the Alternating Minimal Energy (AMEn) algorithm \cite{amen1}. The AMEn algorithm is a combination of the single-block Density Matrix Renormalization Group (DMRG) \cite{DMRG} and the steepest descent \cite{iterative_methods} algorithms. 
At each step, the local system of equations is solved, like in DMRG, but also a corresponding TT-core of the residual $\mathbf{z} = \mathcal{A} \mathbf{x} - \mathbf{b}$ is efficiently added. 
This allows one to vary the ranks of the solution and get rid of the stagnation in the local minima that are inherent in DMRG. 
The AMEn algorithm has complexity $O(dnr(\mathbf{x})^3r(\mathcal{A}) + dnr(\mathbf{x})^2r(\mathcal{A})^2)$.

Finally, the algorithm that was not included in Table~\ref{table:operations} is TT-cross approximation \cite{tt_cross}. 
It enables one to approximate a wide range of functions in the TT format with only $O(dnr^2)$ function evaluations. 
We use this algorithm to represent the Jacobian of the domain transformation and initial conditions in the TT format.

\section{TENSOR-TRAIN FINITE ELEMENT METHOD ({\tt TetraFEM})}\label{TetraFEM}
In this section, we develop our proposed method of using the TT decomposition for the solution of PDEs. 
We use a finite element formulation because it is easier to handle complex areas and domain meshing.

\subsection{Finite element discretization of PDEs}\label{basic}

\begin{figure}
    \centering
    \includegraphics[width=1\linewidth]{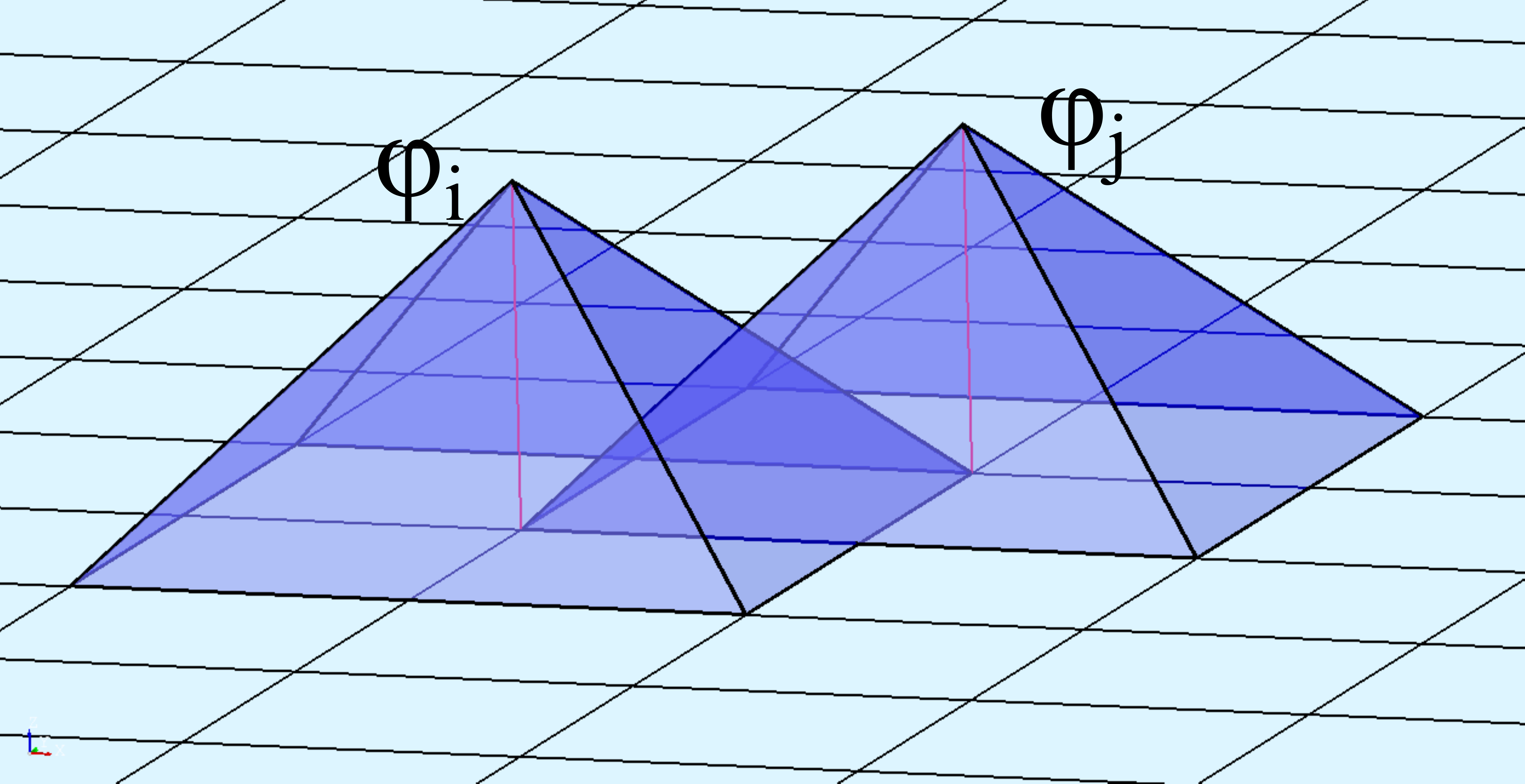}
    \caption{An example of linear basis functions for quadrilateral finite elements in the 2D case. 
    Each function equals one on its node and zero on every other node, which allows the coefficients $\Bar{u}$ in Eq.~\eqref{u} to be interpreted as values of the target function at the nodes. 
    Also, each function is non-zero only in a small area, which results in the sparseness of the matrices in Eq.~\eqref{eq:matrices} and Eq.~\eqref{eq:matrices_Dxy}.}
    \label{fig:base}
\end{figure}

The main idea of the finite element method (FEM) is to approximate a target function, $u(x,y)$, as a sum of $N$ known basis functions $\{\varphi_i(x,y)\}_{i=1}^{N}$ each multiplied by coefficients $\{\bar u_i\}_{i=1}^{N}$,
\begin{equation}\label{u}
    u(x, y) \approx u_N(x,y):=\sum_{i=1}^{N} \bar u_i \varphi_i(x, y).
\end{equation}
These basis functions are non-zero only on a small part of the domain, are differentiable and are of the same type (such as piecewise polynomial). 
Usually, these functions are linked with the discretization mesh, for example, 
$$
\varphi_i(x, y) = \left\{\begin{array}{rl}
1, & \mbox{$(x,y)$ is the $i$-th mesh node}, \\ 
0, & \mbox{$(x,y)$ is the $j$-th mesh node with $j\neq i$,}\\
& \mbox{polynomial$(x,y)$ between mesh nodes.}
\end{array}\right.
$$
In this paper, we use basis functions that are linear between the mesh nodes, as shown in Fig.~\ref{fig:base}.

Partial derivatives of the discrete function follow trivially from Eq.~\eqref{u}:
\begin{gather*}
    \nabla u(x, y) = \sum_{i=1}^{N} \bar u_i \nabla \varphi_i(x, y).
\end{gather*}

Let us consider a model PDE problem 
\begin{align*}
    -\Delta u & = f & \mbox{on a domain } D\subset\mathbb{R}^2, \\
    u & = 0 & \mbox{on the boundary } \partial D,
\end{align*}
where $f\in L^2(D)$ is a forcing term.
The solution $u \in H^1_0(D)$ can be sought in the weak formulation
\begin{equation}\label{integral}
    \iint_D (\nabla u(x, y), \nabla \phi(x, y)) dx dy = \iint_D f(x, y) \phi(x, y) dx dy,
\end{equation}
for all $\phi \in H^1_0(D)$.
After substituting Eq.~\eqref{u} into Eq.~\eqref{integral} and since the basis functions are known, they can be integrated, which results in a linear system of equations:
\begin{equation}\label{syst}
     \mathbf{S} \Bar{u} = \mathbf{M} \Bar{f},
\end{equation}
where $\bar f = \{\bar f_i\}_{i=1}^{N}$ is a vector of coefficients of $f(x,y)$ in the finite element basis and 
$\mathbf{S},  \mathbf{M} \in \mathbb{R}^{N \times N}$ are called the \textit{stiffness} and \textit{mass} matrices, respectively.
They are populated with elements: 
\begin{gather}\label{eq:matrices}
    \mathbf{S}_{ij} = \iint (\nabla \varphi_i, \nabla \varphi_j) dx dy, \quad 
    \mathbf{M}_{ij} = \iint  \varphi_i \varphi_j dx dy.
\end{gather}
Matrices discretizing the first derivatives can be computed similarly:
\begin{gather}\label{eq:matrices_Dxy}
(\mathbf{D_{\mu}})_{ij} = \iint (\varphi_i, \frac{\partial \varphi_j}{\partial \mu}) dx dy, \quad \mu \in \{x,y\}.
\end{gather}
These matrices are sparse, which follows from the fact that each basis function is non-zero only in a small area, so the integrals above are also non-zero only for the neighboring basis functions.
This allows us to solve Eq.~\eqref{syst} efficiently, for example, using iterative methods.
Moreover, as we will see, such matrices are effectively represented in TT format.

%




\subsection{Quantized Tensor Train}

\begin{figure}
        \centering
        \includegraphics[width = 1 \linewidth]{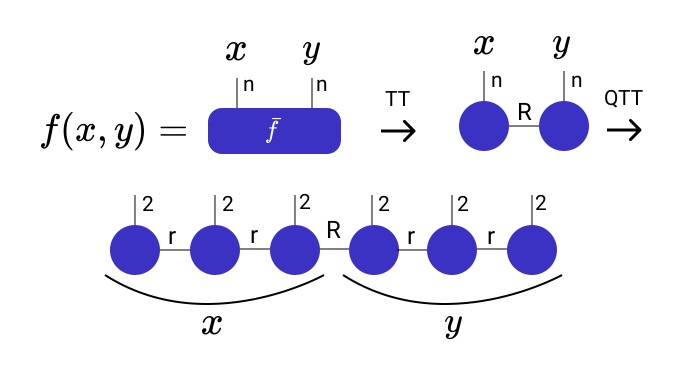}
        \caption{We discretize a two-dimensional function $f(x, y)$ on a uniform grid and obtain the corresponding tensor $\bar{f}$. 
        We represent this tensor in a Tensor Train (TT) format and then further compress it into a Quantized Tensor Train (QTT) format, where each visible index has dimension 2.}
        \label{fig:qtt_canonical}
\end{figure}

The functions we consider in this paper seemingly depend on only two variables, $x$ and $y$.
To compress coefficient vectors, such as $\bar u$ or $\bar f$, we use the so-called Quantized Tensor Train (QTT) format \cite{QTT}. 
It is obtained by splitting the index $i=1,\ldots,N$ into artificial indices of smaller range,
$$
i = 1 + (i_1-1) + 2 (i_2-1) + \cdots + 2^{d-1}(i_d-1),
$$
where $d=\log_2 N$, such that $i_k \in \{1,2\}$ and $k=1,\ldots,d$.
A coefficient vector, for example, $\bar f$, can now be reshaped into a tensor with elements $\bar f(i_1,\ldots,i_d)$.
As shown in Fig.~\ref{fig:qtt_canonical}, this tensor can be approximated in the TT format as before but the fact that the indices $i_1,\ldots,i_d$ ``quantize" the original index $i$ is referred to by naming the resulting representation the QTT format of $\bar f$.
The utility of the QTT format stems from the fact that the FEM expansion coefficients of many elementary functions (such as trigonometric, exponential, polynomial and rational functions) admit rapidly converging, or even exact, QTT decompositions \cite{QTT_functions}.




\begin{figure*}
    \centering
    \includegraphics[width=1\linewidth]{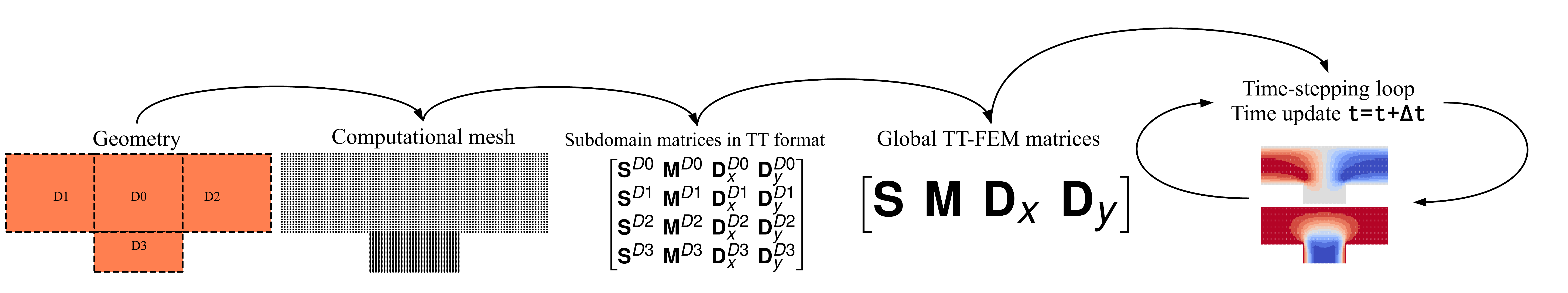}
    \caption{Scheme of the full simulation process via {\tt TetraFEM}. 
    At first, the geometry is specified as a set of adjacent subdomains (quadrilaterals with analytically-defined boundaries). 
    Then, each of the subdomains is transformed into a rectangle and the mesh is generated.
    After that, the Jacobian of the transformation is sampled on the mesh and local finite element matrices (in each of the subdomains) are assembled in the Tensor Train format. 
    The global stiffness, mass and partial derivatives matrices are completed as a concatenation of local ones.
    Finally, after setting the initial and boundary conditions, the time-stepping can be performed as a sequence of TT-operations.}
    \label{fig:scheme}
\end{figure*}


In addition, it is experimentally shown that the finite element matrices have low QTT ranks, and therefore are well-compressible (Fig.~\ref{fig:memory}) \cite{markeeva}. This means that all matrix-vector operations from Table~\ref{table:operations} will have lower computational complexity than operations with full (even sparse) matrices.

\begin{figure}[H]
    \centering
    \includegraphics[width=1\linewidth]{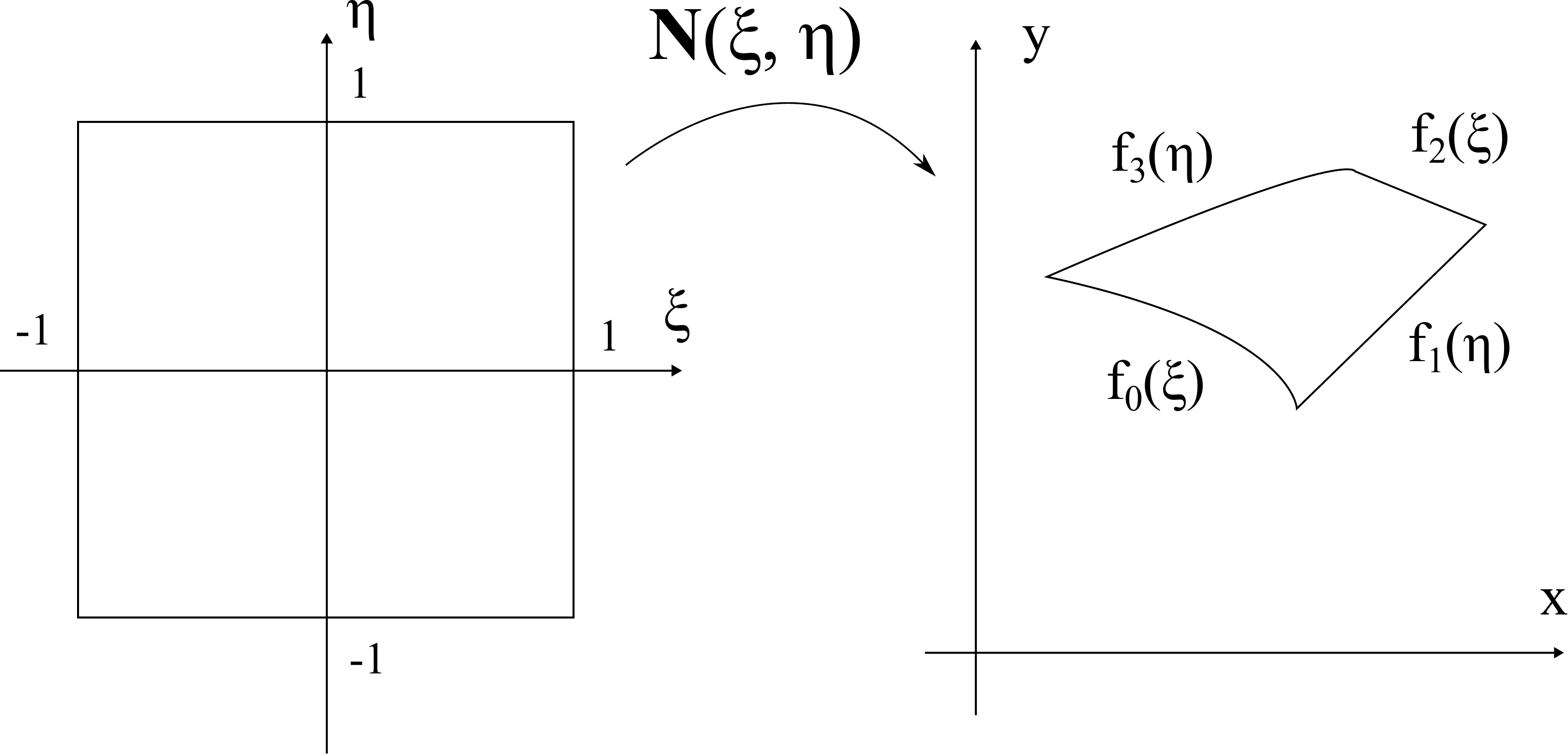}
    \caption{The projection of a reference square onto the domain of Eq.~\eqref{eq:transformer}. 
    The curved sides of the domain are described with four differentiable analytic functions.}
    \label{fig:quad}
\end{figure}

\subsection{Domain geometry and mesh}

We consider the domain $D$ which consists of $S$
arbitrary curvilinear quadrilaterals that are connected by their sides as shown in Fig.~\ref{fig:domain}. Each subdomain is specified by a set of four parametrized smooth curves: 
\begin{gather*}
    \mathbf{f}_k: t \longrightarrow (x, y), \quad t \in [-1, 1], \quad k=1, \dots 4, \\
    \mathbf{f}_0(-1)=\mathbf{f}_3(-1), \quad \mathbf{f}_1(-1)=\mathbf{f}_0(1), \\
    \mathbf{f}_2(1)=\mathbf{f}_1(1), \quad \mathbf{f}_3(1)=\mathbf{f}_2(-1).
\end{gather*}
The last four equations define the directions of parametrizations and guarantee that the sides of a quad coincide at the corners. After that, the differentiable mapping from a reference $[-1,1] \times [-1,1]$ square to a quad can be constructed:
\begin{multline}\label{eq:transformer}
    \mathbf{N}(\xi, \eta) = -\frac{(1 - \xi)(1 - \eta)}{4}\mathbf{f}_0(-1) - \\ 
    \frac{(1 + \xi)(1 - \eta)}{4}\mathbf{f}_1(-1) - \frac{(1 + \xi)(1 + \eta)}{4}\mathbf{f}_2(1) - \\ 
    \frac{(1 - \xi)(1 + \eta)}{4}\mathbf{f}_3(1) + \frac{(1 - \eta)}{2}\mathbf{f}_0(\xi) + \\
    \frac{(1 + \xi)}{2}\mathbf{f}_1(\eta) + \frac{(1 + \eta)}{2}\mathbf{f}_2(\xi) + \frac{(1 - \xi)}{2}\mathbf{f}_3(\eta).
\end{multline}
This technique is also known as transfinite interpolation \cite{GORDON1982171}.
The Jacobian of this transformation can be computed as follows:
\begin{gather}\label{eq:jacobian}
    J = \begin{bmatrix} J_{11} & J_{12} \\ J_{21} & J_{22} \end{bmatrix}
    =
    \begin{bmatrix}
        \frac{\partial x}{\partial \xi} & \frac{\partial x}{\partial \eta} \\
        \frac{\partial y}{\partial \xi} & \frac{\partial y}{\partial \eta}
    \end{bmatrix}
    =
    \begin{bmatrix}
        \frac{\partial \mathbf{N}(\xi, \eta)}{\partial \xi} & \frac{\partial \mathbf{N}(\xi, \eta)}{\partial \eta}
    \end{bmatrix}.
\end{gather}

To set the finite element mesh (i.e., the positions of the basis functions), a uniform Cartesian mesh of size $2^{d/2} \times 2^{d/2}$ is introduced on the reference square and then projected onto each quadrilateral with Eq.~\eqref{eq:transformer} as shown in Fig.~\ref{fig:quad}. 
Using this and also the analytic formula for the Jacobian (Eq.~\eqref{eq:jacobian}), finite element matrices from Eq.~\eqref{eq:matrices} and Eq.~\eqref{eq:matrices_Dxy} are constructed for each of the $S$ subdomains. 
So Eq.~\eqref{eq:matrices} becomes 
\begin{equation}\label{jacobian}
    \mathbf{S}_{ij} = \iint\limits_{[-1, 1] \times [-1, 1]} \Big((J^T)^{-1}\nabla \Phi_i, (J^T)^{-1}\nabla \Phi_j\Big) |J| dx dy,
\end{equation} where $\Phi$ is a $Q1$ basis function on the reference square and $J$ is the Jacobian. The numerical integration is performed by standard $2 \times 2$ Gauss–Legendre quadrature \cite{fem}. The matrices $\mathbf{M}, \mathbf{D_x}$ and $\mathbf{D_y}$ are assembled analogously. To represent the values of the Jacobian for the arbitrary $\xi$ and $\eta$, the cross-approximation approach is utilized \cite{tt_cross}.

This results in the adjacent subdomain meshes having common nodes on the inner boundaries which can be seen in Fig.~\ref{fig:mesh}.
The coefficient vectors $\bar f_1,\ldots,\bar f_S$, corresponding to different subdomains, are enumerated by an extra index $s=1,\ldots,S$. 
This allows us to store them all in a tensor with elements $\bar f(i_1,\ldots,i_d,s)$ and approximate it in a $(d+1)$-dimensional QTT format.




\subsection{Assembly of global FEM matrices}



During the assembly, \textit{local} finite element matrices are generated on each of the $S$ quadrilateral subdomains. Then, the \textit{global} matrices are obtained by so-called \textit{subdomain stitching}. Let's denote one of the $\mathbf{M}, \mathbf{S}, \mathbf{D_x}, \mathbf{D_y}$ matrices as $\mathbf{A}$, because the stitching procedure is identical for every type of matrix.

The idea of stitching is to unite the local basis functions on the borders of the adjoined subdomains into continuous functions in the global block matrix by handling the indices during the construction of blocks.

The diagonal block of the global $\mathbf{A}$ is defined as follows:
\begin{gather*}
\mathbf{A}^{k k} = \mathbf{A}^k + c \mathbf{P}^k,
\end{gather*}
where $\mathbf{A}^k$ is the local matrix of the $k$-th subdomain, $\mathbf{P}^k$ is the diagonal matrix with ones on the positions corresponding to the indices of the repeating basis functions and $c$ is the maximum absolute value across all the main diagonals of $\mathbf{A}^k$ for every $k$.

The construction of non-diagonal blocks is a little more complicated:
\begin{gather*}
\mathbf{A}^{k l} = \mathbf{P}^{k l} \mathbf{A}^l - c \mathbf{P}^{k l}.
\end{gather*}
Here, $\mathbf{P}^{k l}_{ij} = 1$ if the $i$-th basis function of the $k$-th subdomain globally coincides with the $j$-th function of $l$-th subdomain and is zero otherwise. The values are being ``pulled" from the neighbor subdomain matrix.

The final assembly of the global matrix is performed via Kronecker product in the Tensor Train format:
\begin{gather*}
\mathbf{A} = \sum_{k,l = 1}^{S} \mathbf{A}^{kl} \otimes \mathbf{I}^{kl},
\end{gather*}
where $\mathbf{I}^{kl}$ is the $S \times S$ matrix with the only non-zero element $\mathbf{I}^{kl}_{kl} = 1$.



\section{GOVERNING EQUATIONS AND NUMERICAL SCHEME}\label{equations}

\subsection{General numerical scheme}

To model the flow, the incompressible Navier-Stokes equations are considered:

\begin{gather*} 
        \frac{\partial \boldsymbol{u}}{\partial t} + \boldsymbol{u} \cdot \nabla \boldsymbol{u} = - \frac{1}{\rho} \nabla p + \nu \nabla^{2} \boldsymbol{u}, \\
        \nabla \cdot \boldsymbol{u} = 0,
\end{gather*}
where $\boldsymbol{u} = [u \quad v]^T$ stands for the X- and Y-components of the velocity, $\rho$ is the constant fluid density, $p$ is the pressure and $\nu$ is the kinematic viscosity. The first equation is called the momentum equation and the second is the continuity equation.

The possible boundary conditions include:

\begin{enumerate}
    \item \emph{Inlet.} $$ \boldsymbol{u} = \boldsymbol{u}_{in} $$
    This boundary condition implies that the inflow is known from the problem statement and therefore, the velocity on the boundary remains constant for each time step. The velocity can be constant across the boundary or parabolic, which is used to eliminate unrealistic gradients at the corners: $$u_{in}(0, y) = u_0 ( 1 - (1 - 2y)^2 ), \quad v_{in}(0,y) = 0.$$
    \item \emph{Outlet.} $$\nu\frac{\partial \boldsymbol{u}}{\partial \boldsymbol{n}} - \boldsymbol{n}p = \boldsymbol{0}$$
    The outflow means that the fluid is leaving the domain unconstrained.
    This approach is called the ``do-nothing boundary condition" and is widely used in computational fluid dynamics. 
    
    \item \emph{No-slip.} $$ \boldsymbol{u} = \boldsymbol{0} $$
    This boundary condition means that the walls are stationary and ``sticky." 
    In some cases, the wall can be considered ``moving" and then the velocity on this wall has some fixed non-zero value.
    \item \emph{Natural.} $$ \frac{\partial p}{\partial \boldsymbol{n}} = 0$$
    The natural boundary condition is used when solving Poisson's equation for the pressure. 
    After inspecting the momentum equation, one can notice that a constant can be added to the pressure without affecting the result.
\end{enumerate}

To impose the boundary conditions, a $\mathbf{Mask}$ matrix is used:
\begin{equation*}
\mathbf{Mask}_{i,j} =
  \begin{cases}
    1  &  \text{if } i=j \text{ and $\varphi_i$ is interior or outlet} \\
    0  &  \text{else}
  \end{cases}
\end{equation*}

The $\mathbf{Mask}$ is diagonal so that after multiplying the solution by the mask, the essential boundary values become zero. 
Subsequently, any values can be inscribed by adding a vector with boundary values (e.g., $u_{in}$ and $v_{in}$) at the elements corresponding to the essential boundary nodes, and zeros elsewhere.


For the numerical solution, we chose Chorin's projection method \cite{chorin} -- the temporal discretization is explicit Euler: $$ \frac{ \boldsymbol{u}^{n+1} - \boldsymbol{u}^{n} }{\Delta t} = R(t^n),$$ where $t^n = n \Delta t$ is the $n$-th point in time and $R(t)$ is the right hand side of the momentum equation.

The update of the solution on each time step is performed in three substeps:

\begin{enumerate}
    \item \emph{The predictor step.} The intermediate velocity, $\boldsymbol{u}^{*}$, is computed without the pressure taken into account: $$ \frac{\boldsymbol{u}^{*} - \boldsymbol{u}^n}{\Delta t} = - \boldsymbol{u}^{n} \cdot \nabla \boldsymbol{u}^{n} + \nu \nabla^2 \boldsymbol{u}^{n} $$
    \item \emph{The pressure step.} Here, Poisson's equation is solved for the pressure. $$ \nabla^2 p^{n+1} = \frac{\rho}{\Delta t} \nabla \cdot \boldsymbol{u}^{*} $$
    \item \emph{The corrector step.} In this step, the pressure term is finally added to the equation. $$ \frac{\boldsymbol{u}^{n+1} - \boldsymbol{u}^{*}}{\Delta t} = -\frac{1}{\rho} \nabla p^{n+1}$$
\end{enumerate}

It can be observed that the sum of the predictor and corrector steps results in the momentum equation. 
For the convergence of the scheme, it is necessary to ensure two conditions on the time step. 
The first one is the Courant–Friedrichs–Lewy (CFL) condition for the convection term $\boldsymbol{u} \cdot \nabla \boldsymbol{u}$: $$ \Delta t_1 < \frac{ C_1 \Delta x }{ |u|_{\max} }, $$ where $C_1 < 1$ is the CFL number. 
For the diffusion term, $\nu \nabla^2 \boldsymbol{u}$, the condition is $$ \Delta t_2 < \frac{\Delta x^2}{\nu}. $$
So the time step $\Delta t$ is chosen such that $\Delta t \leq \min(\Delta t_1, \Delta t_2)$.

\begin{figure*}
\centering
\begin{subfigure}{0.48\textwidth}
   \includegraphics[width=1\linewidth]{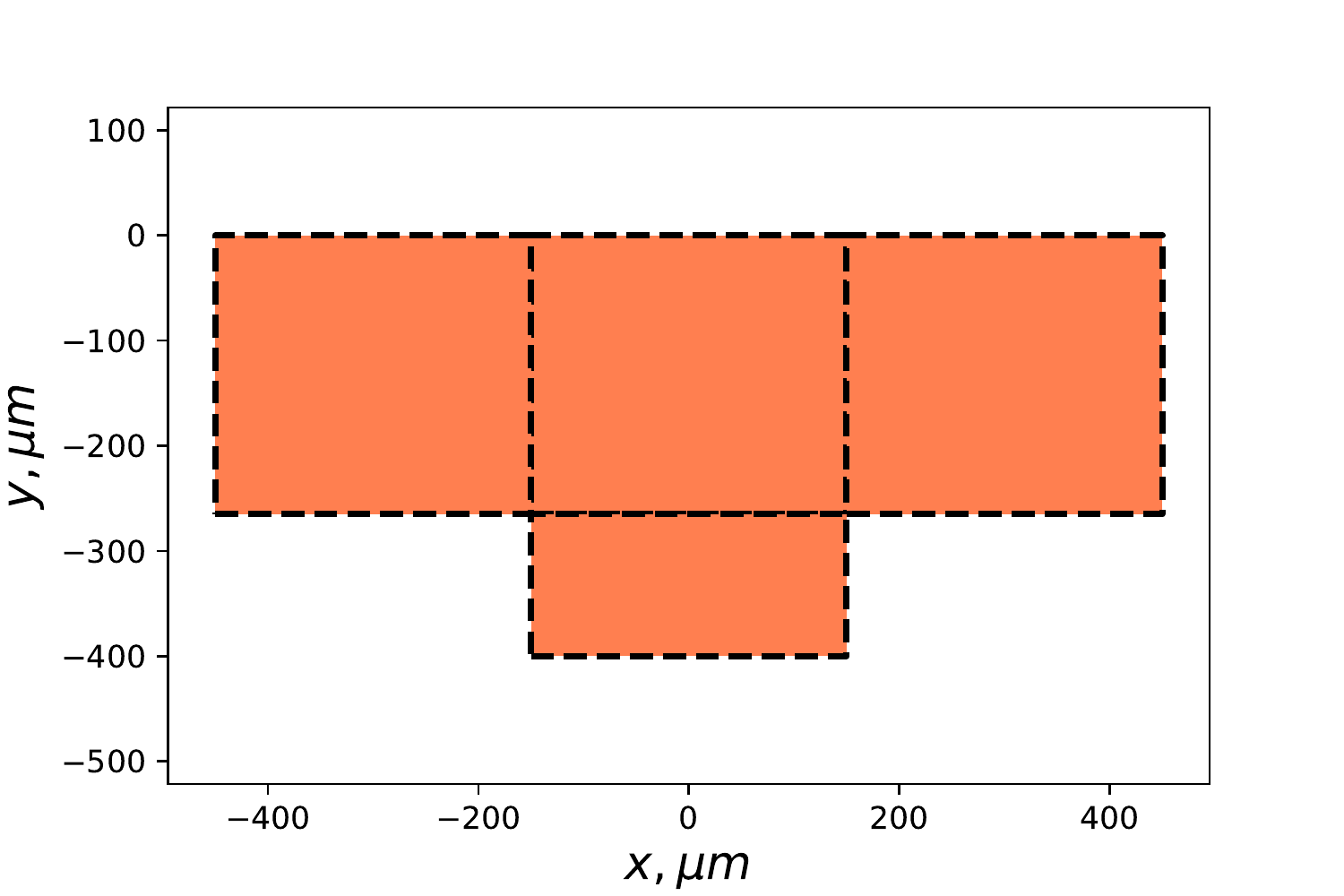}
   \caption{The T-mixer geometry, representing an intersection of two straight microchannels.}
   \label{fig:domain} 
\end{subfigure}
\hfill
\begin{subfigure}{0.48\textwidth}
   \includegraphics[width=1\linewidth]{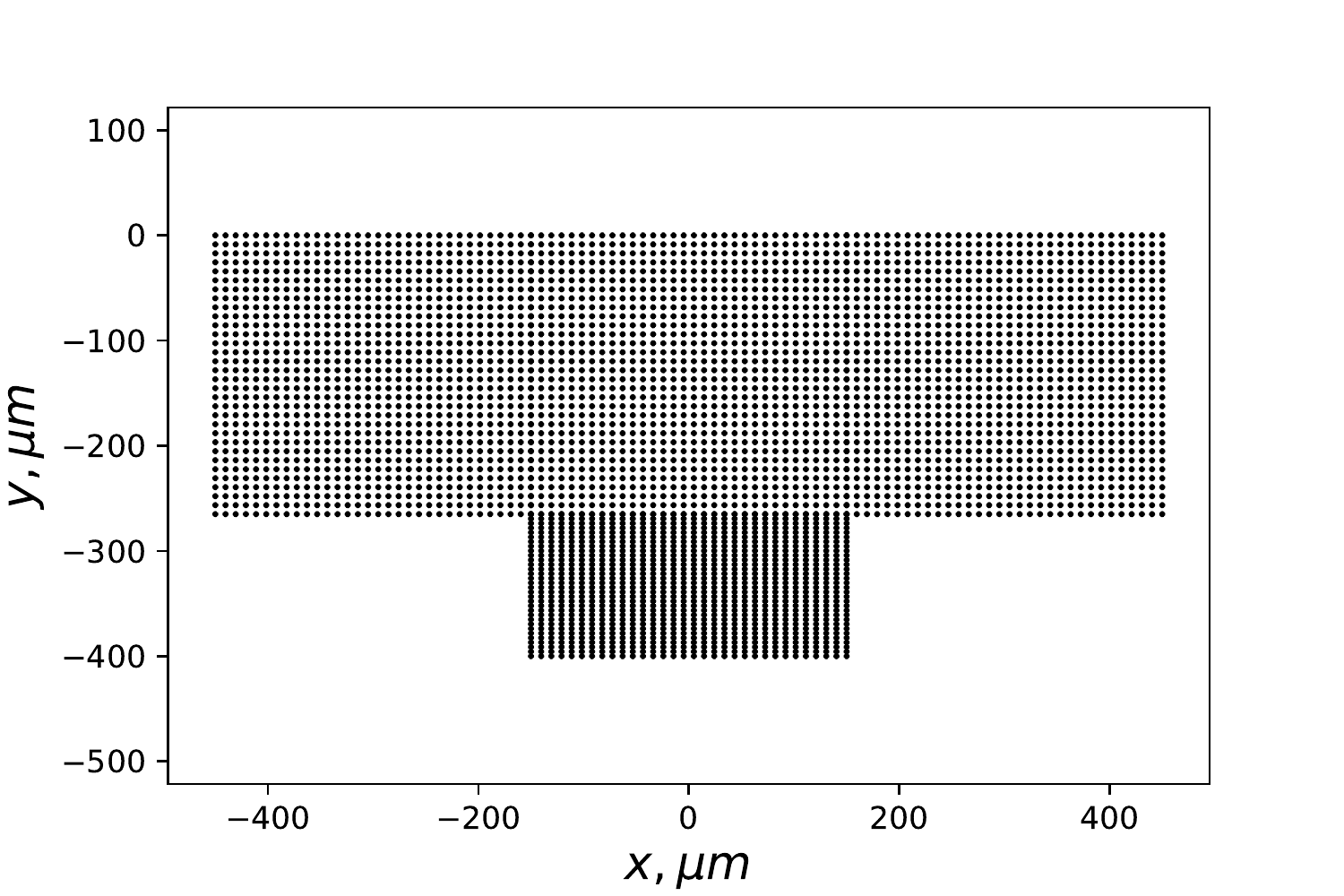}
   \caption{Computational mesh (centers of basis functions).}
   \label{fig:mesh}
\end{subfigure}

\caption{An example of a computational domain. 
(a) depicts the geometry specified as a set of connected quadrilaterals and (b) shows the domain after assigning the mesh for each subdomain.}
\end{figure*}





\subsection{Realization of numerical scheme using Tensor Train operations}

After assembling all the matrices, the scheme can be realized as shown in Algorithm~\ref{alg2}.

From this point on, we will be considering the X and Y components of the velocity $\mathbf{u} = [u, v]$ from Section~\ref{equations} separately as two TT-vectors of discrete coefficients,
$u(i_1,\ldots,i_d) = \bar u_i$ (as defined in Eq.~\eqref{u}), and $v(i_1,\ldots,i_d) = \bar v_i$;
the pressure $p$ is represented analogously. 

As long as only a limited amount of operations are needed for the basic method (such as the element-wise sum and product of two vectors, matrix-vector multiplication and the solving of a linear system), which are all realized in tensor-train format (Table~\ref{table:operations}) -- after assembling the TT-finite element matrices, the algorithm could be rewritten as a sequence of operations over TT-vectors. 
The notation is as follows: $u$ and $v$ stand for the x- and y-components of the velocity; $\varepsilon$ is the relative error of the TT-{\tt rounding}; TT-matrices are shown in bold; TT-vectors are not in bold.

The solution of linear systems is performed with the AMEn algorithm \cite{amen1}, which is implemented in a function $amen\_solve$ in the \texttt{ttpy} package \cite{ttpy}.

\begin{algorithm}[H]
\caption{Predictor-corrector time stepping}\label{alg2}
\begin{algorithmic}[1]
\State $u = u_{in}.copy()$
\State $v = v_{in}.copy()$
\For{$it = 1, nt$}
    \State $\triangleright$ {\it Compute numerical derivatives}
    \State $Su = solve(\mathbf{M}, matvec(\mathbf{S}, u)).round(\varepsilon)$
    \State $Sv = solve(\mathbf{M}, matvec(\mathbf{S}, v)).round(\varepsilon)$
    \State $Dxu = solve(\mathbf{M}, matvec(\mathbf{D_x}, u)).round(\varepsilon)$
    \State $Dxv = solve(\mathbf{M}, matvec(\mathbf{D_x}, v)).round(\varepsilon)$
    \State $Dyu = solve(\mathbf{M}, matvec(\mathbf{D_y}, u)).round(\varepsilon)$
    \State $Dyv = solve(\mathbf{M}, matvec(\mathbf{D_y}, v)).round(\varepsilon)$
    \State $\triangleright$ {\it Predictor step}
    \State $u^* = u + \Delta t (\nu Su - u \odot Dxu - v \odot Dyu)).round(\varepsilon)$
    \State $v^* = v + \Delta t (\nu Sv - u \odot Dxv - v \odot Dyv)).round(\varepsilon)$
    \State $\triangleright$ {\it Pressure-Poisson equation}
    \State $Dxu = matvec(\mathbf{D_x}, u^*$)
    \State $Dyv = matvec(\mathbf{D_y}, v^*$)
    \State $f = \frac{\rho}{\Delta t}(Dxu + Dyv).round(\varepsilon)$
    \State $p = solve(\mathbf{S}, f)$
    \State $\triangleright$ {\it Corrector step}
    \State $Dxp = solve(\mathbf{M}, matvec(\mathbf{D_x}, p)).round(\varepsilon)$
    \State $Dyp = solve(\mathbf{M}, matvec(\mathbf{D_y}, p)).round(\varepsilon)$
    \State $u = u^* - \frac{\Delta t}{\rho} Dxp$
    \State $v = v^* - \frac{\Delta t}{\rho} Dyp$
    \State $\triangleright$ {\it Apply boundary conditions and update the values}
    \State $u = matvec(\mathbf{Mask}, u) + u_{in}$
    \State $v = matvec(\mathbf{Mask}, v) + v_{in}$
\EndFor

\end{algorithmic}
\end{algorithm}


The frequent TT-rounding is used to prevent uncontrolled growth of the ranks and to keep the operation asymptotics low (see Table~\ref{table:operations}). 
The fine tuning of $\varepsilon$ allows us to control the accuracy of the solution and establish the optimal trade-off between the error and the computational time.

\section{Numerical experiment -- T-Mixer problem}\label{test}

The main problem that we solve here is a mixing of liquids in a rectangular T-mixer. 
We took the problem statement and the geometry parameters of the T-mixer (T-junction) from Ref.~\cite{tmixer}:

\begin{table}[H]\label{table:results_comparison}
\centering
	\begin{tabular}{|l|l|}
		\hline
 	  {\bf Parameter}         &  {\bf Value~~}  \\ \hline
        Inlet width, \textmu m 	~~ &       $265$            \\ \hline
        Inlet channel length, \textmu m  ~~	 &       $900$            \\ \hline
        Outlet width, \textmu m ~~ &       $300$            \\ \hline
        Outlet channel length, \textmu m ~~	 &       $135$            \\ \hline     
	\end{tabular}
 \end{table}
Here, two flows of liquid are entering the junction by the inlets with such velocities that the flow rates are $4$ ml/h at each channel. 
We consider only the case of a mixture of a single liquid -- pure water at ambient temperature with standard parameters.
The Reynolds number for this problem is estimated to be about $Re \approx 10$.

\begin{figure*}
\centering
\begin{subfigure}{0.48\textwidth}
   \includegraphics[width=1\linewidth]{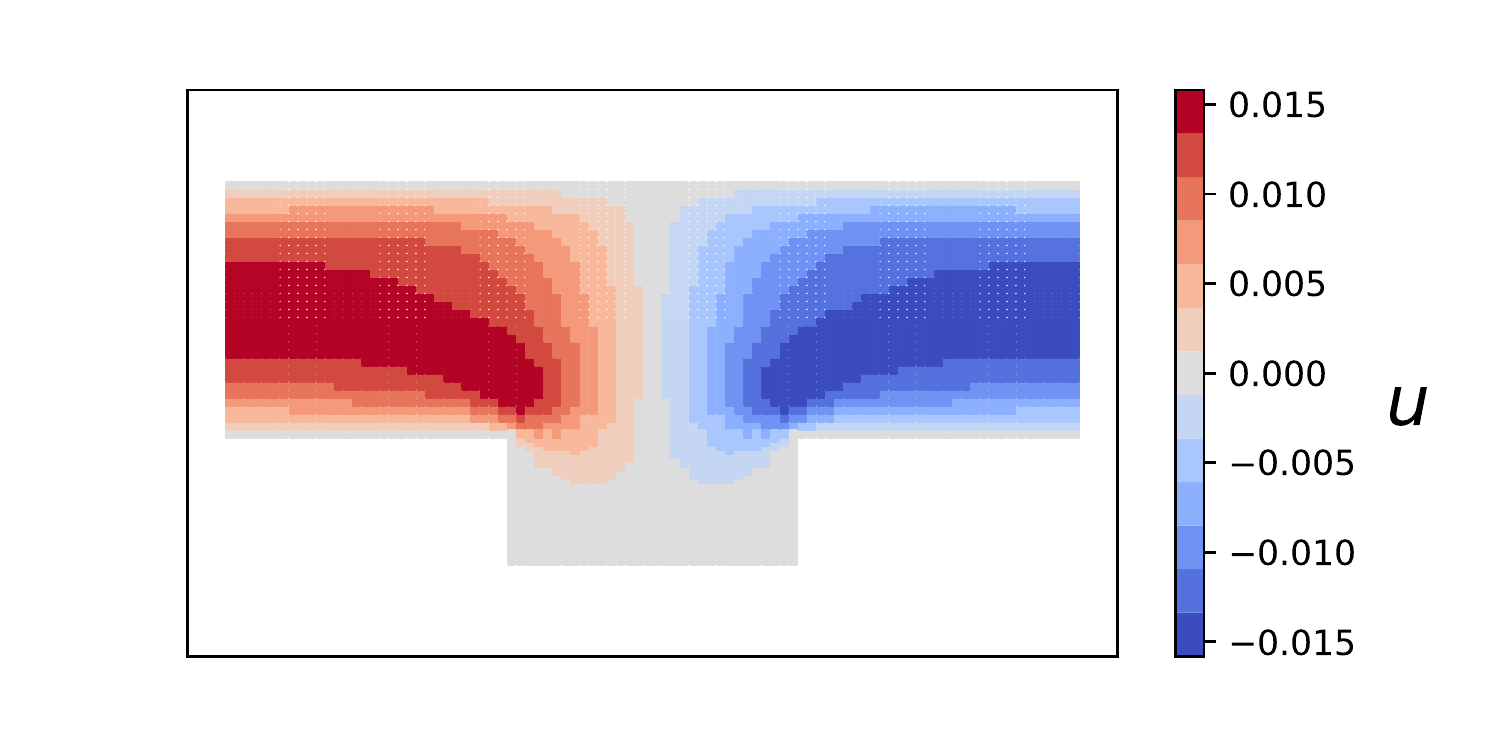}
   \caption{X component of the velocity.}
   \label{fig:u} 
\end{subfigure}
\hfill
\begin{subfigure}{0.48\textwidth}
   \includegraphics[width=1\linewidth]{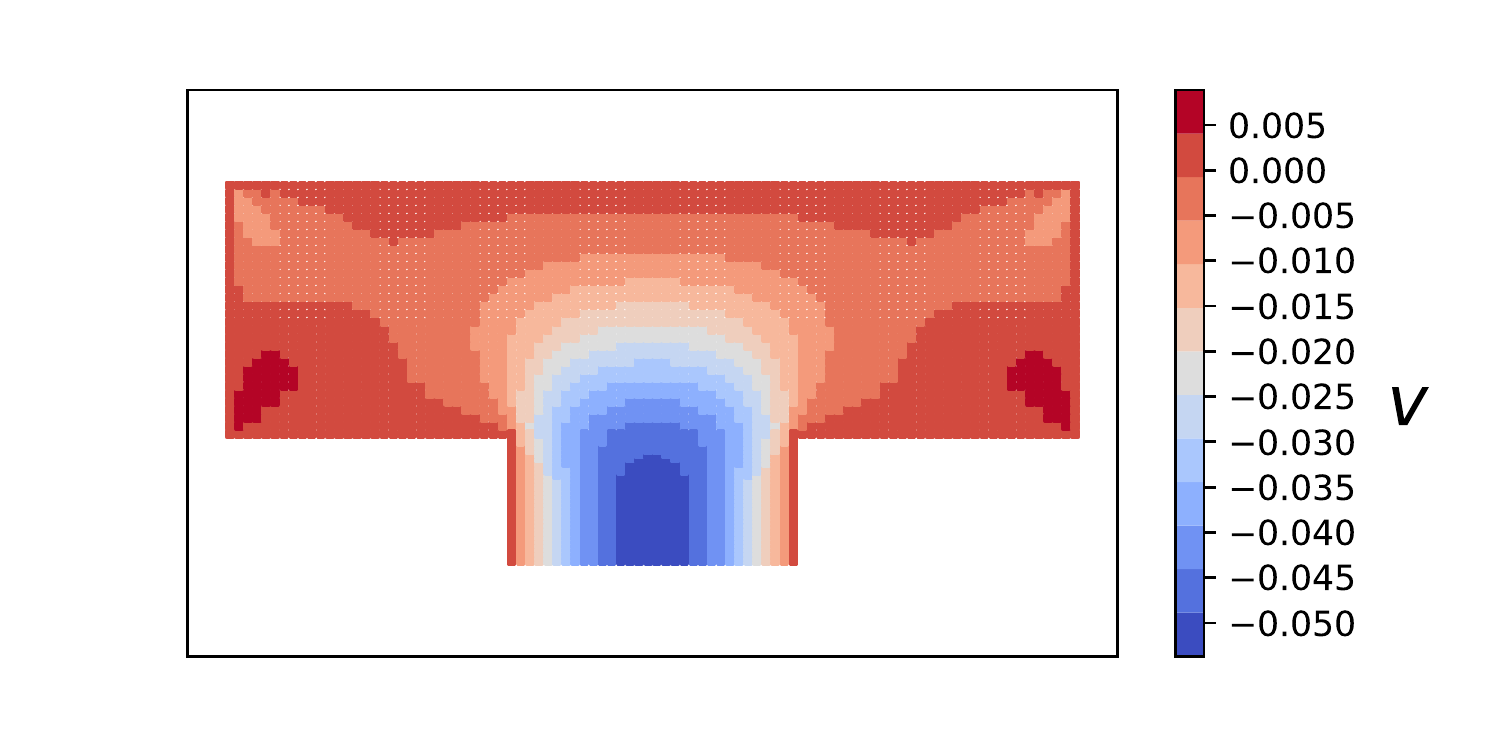}
   \caption{Y component of the velocity.}
   \label{fig:v}
\end{subfigure}

\caption{Values of the established velocities across the domain. The symmetric laminar flow pattern can be observed, which agrees with the physics of the problem. }
\end{figure*}

\begin{figure}
    \centering
    \includegraphics[width=1\linewidth]{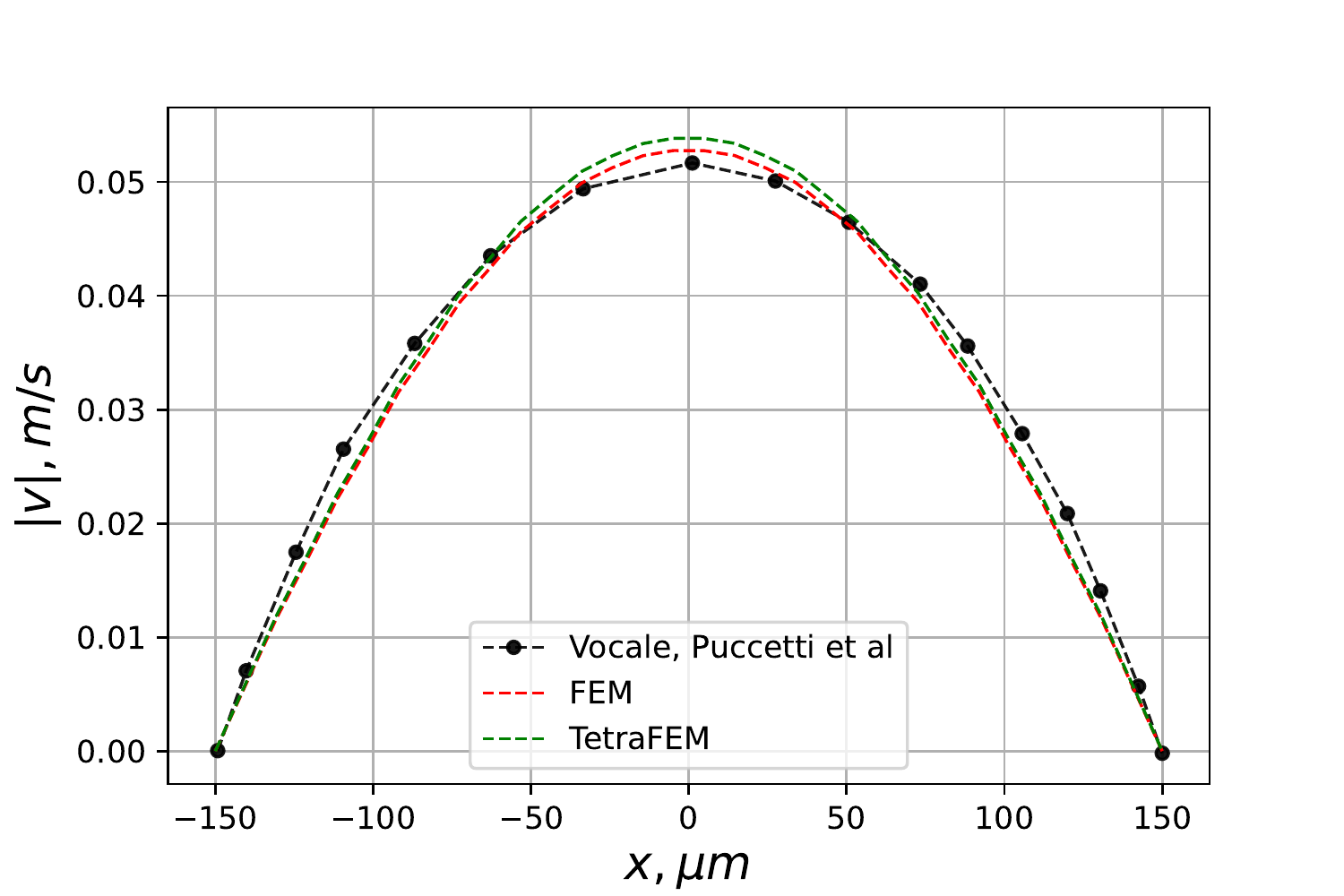}
    \caption{Y-velocity comparison in the fully developed region of the outlet of the T-mixer. 
    The red and green dashed lines represent conventional (FEM) and Tensor Train Finite Element Method ({\tt TetraFEM}) solutions, respectively. 
    The black line refers to the experimental data from Ref.~\cite{tmixer}.}
    \label{fig:1d}
\end{figure}

To solve this problem via {\tt TetraFEM}, we divide the domain into four rectangles: one for the outlet section, two for the inlet sections, and one for the cross-section which are aligned as shown in Fig.~\ref{fig:domain}.
To test the method, we set the rounding tolerance to $\varepsilon = 10^{-3}$ and the discretization parameter to $d = 10$, which means the total number of nodes of the space grid in one subdomain is $2^5 \times 2^5$. 
Figs.~\ref{fig:u} and \ref{fig:v} depict the velocity fields on the outlet in the reached steady state. 
Fig.~\ref{fig:1d} shows a comparison between FEM ({\tt numpy} realization), {\tt TetraFEM} and experimental data \cite{tmixer} for the velocity profiles at the outlet: the curves are almost the same.

\begin{figure}
\centering
\begin{subfigure}[b]{0.49\textwidth}
   \includegraphics[width=1\linewidth]{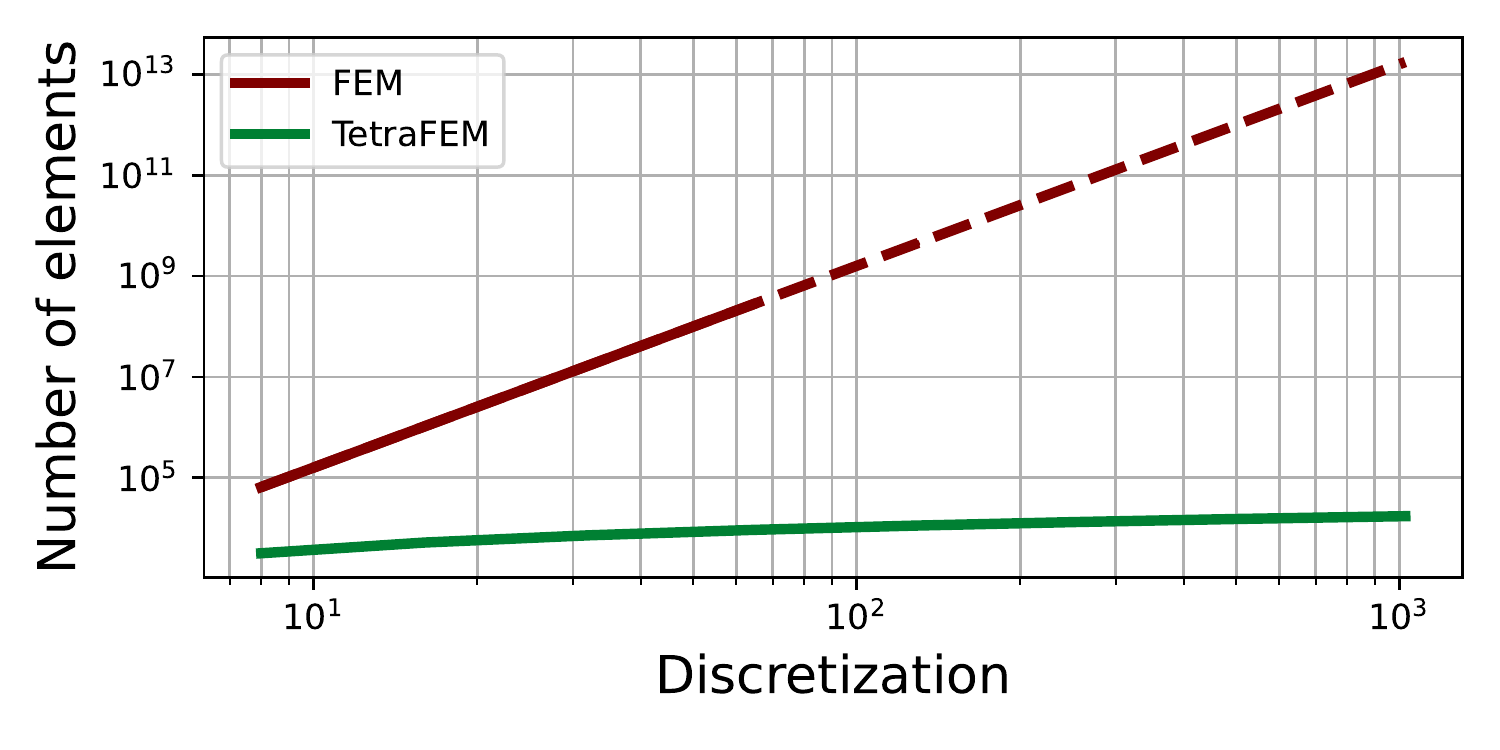}
   \caption{The number of parameters required to store the stiffness matrix $\mathbf{S}$ for the T-mixer geometry. 
   A similar plot is observed for the other FEM matrices. 
   For all discretization values, the TT format allows efficient compression of the matrices. }
   \label{fig:memory} 
\end{subfigure}

\begin{subfigure}[b]{0.49\textwidth}
   \includegraphics[width=1\linewidth]{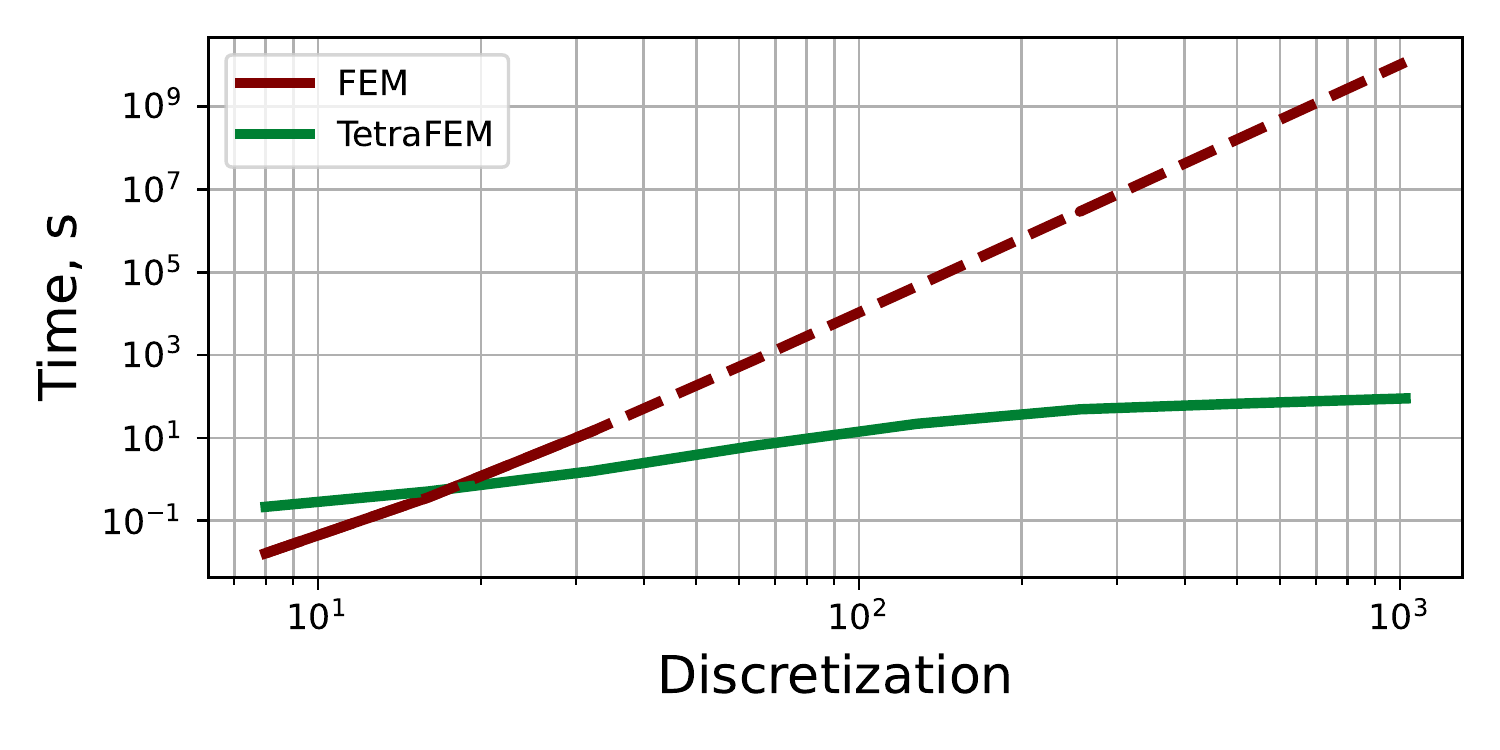}
   \caption{The mean duration of a single time step of the T-mixer flow simulation. 
   On dense meshes, our approach ({\tt TetraFEM}) demonstrates an exponential speed-up compared to the conventional FEM implementation.}
   \label{fig:time}
\end{subfigure}
\caption{
Comparison between the full and TT versions of the Finite Element Method calculation.
The X-axis corresponds to the discretization factor (number of grid points along one axis in a single subdomain). 
(a) depicts the number of parameters required to store the stiffness matrix, (b) compares real times required to perform a single time-step of Algorithm \ref{alg2}. 
When the number of discretization points exceeds $64$, the memory of our machine becomes insufficient to store all the matrices, so the dashed line is obtained by extrapolation.}
\end{figure}

Figs.~\ref{fig:memory} and \ref{fig:time} depict the number of parameters required to keep the stiffness matrix and mean time required to perform a step of the simulation with and without tensor-train format, respectively. 
It can be seen that the stiffness matrix in TT format requires significantly less memory compared to the full format. 
In addition, the runtime comparison demonstrates an exponential speed-up of {\tt TetraFEM} in comparison to the conventional realization when the discretization exceeds $64$ points per axis in a single subdomain.

\begin{figure}
    \centering
    \includegraphics[width=1\linewidth]{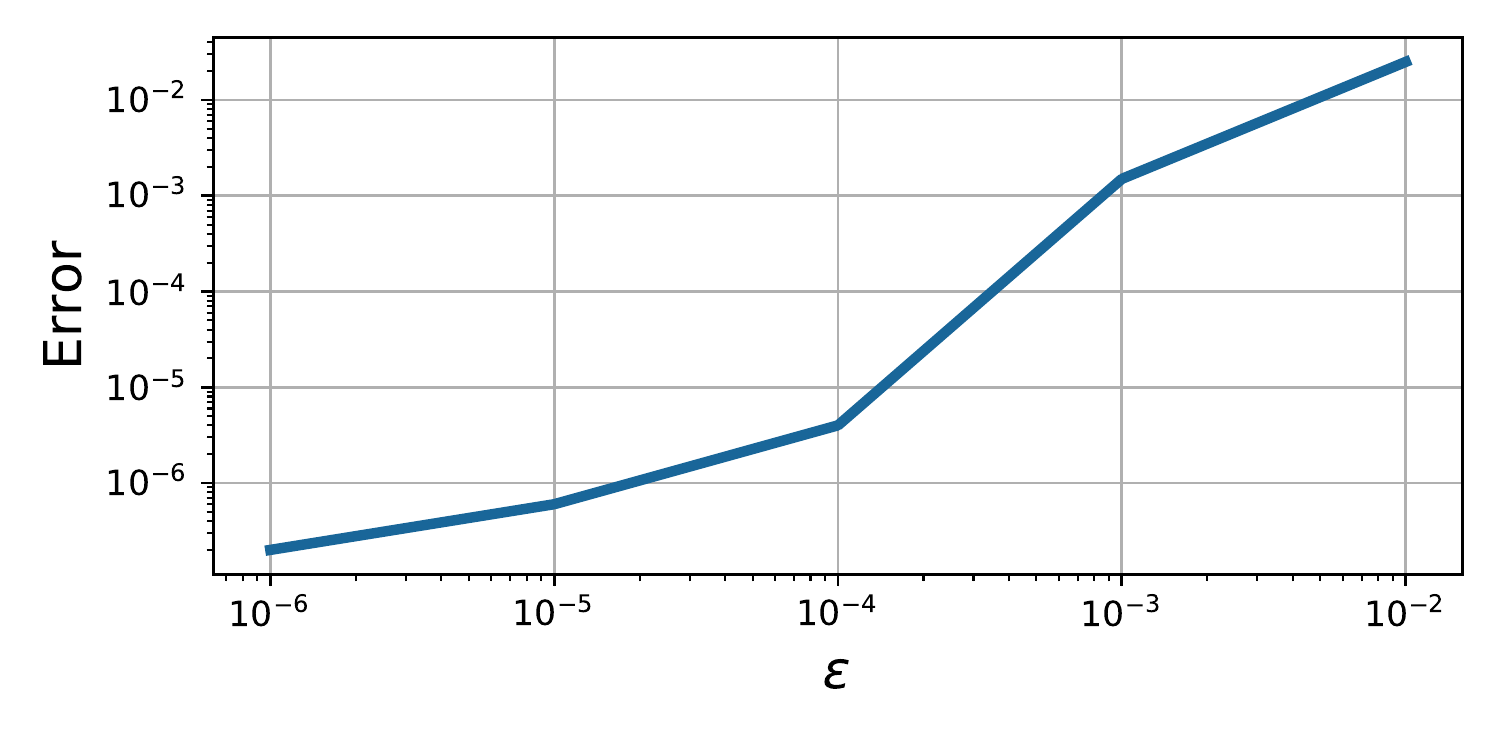}
    \caption{The dependence of the relative error of our implementation ({\tt TetraFEM}) compared to the conventional one (FEM) for various {\tt Rounding }tolerances $\varepsilon$. 
    Here, the mesh discretization is equal to $32$.}
    \label{fig:error}
\end{figure}

We also analyzed the accuracy of the {\tt TetraFEM} solution depending on the TT-rounding tolerance $\varepsilon$ (Fig.~\ref{fig:error}). 
The relative error of {\tt TetraFEM} was calculated with respect to the conventional FEM solution in terms of the Frobenius norm between the X-component of the velocities. 
The results mean that based on the requirements of the method, one can manipulate the TT-rounding error to get more accurate or faster solutions without altering the mesh or other parameters of the method.

We performed all the computations on two Intel Xeon 2.2 GHz CPUs and 12 Gb of RAM.

\begin{figure}
    \centering
    \includegraphics[width=1\linewidth]{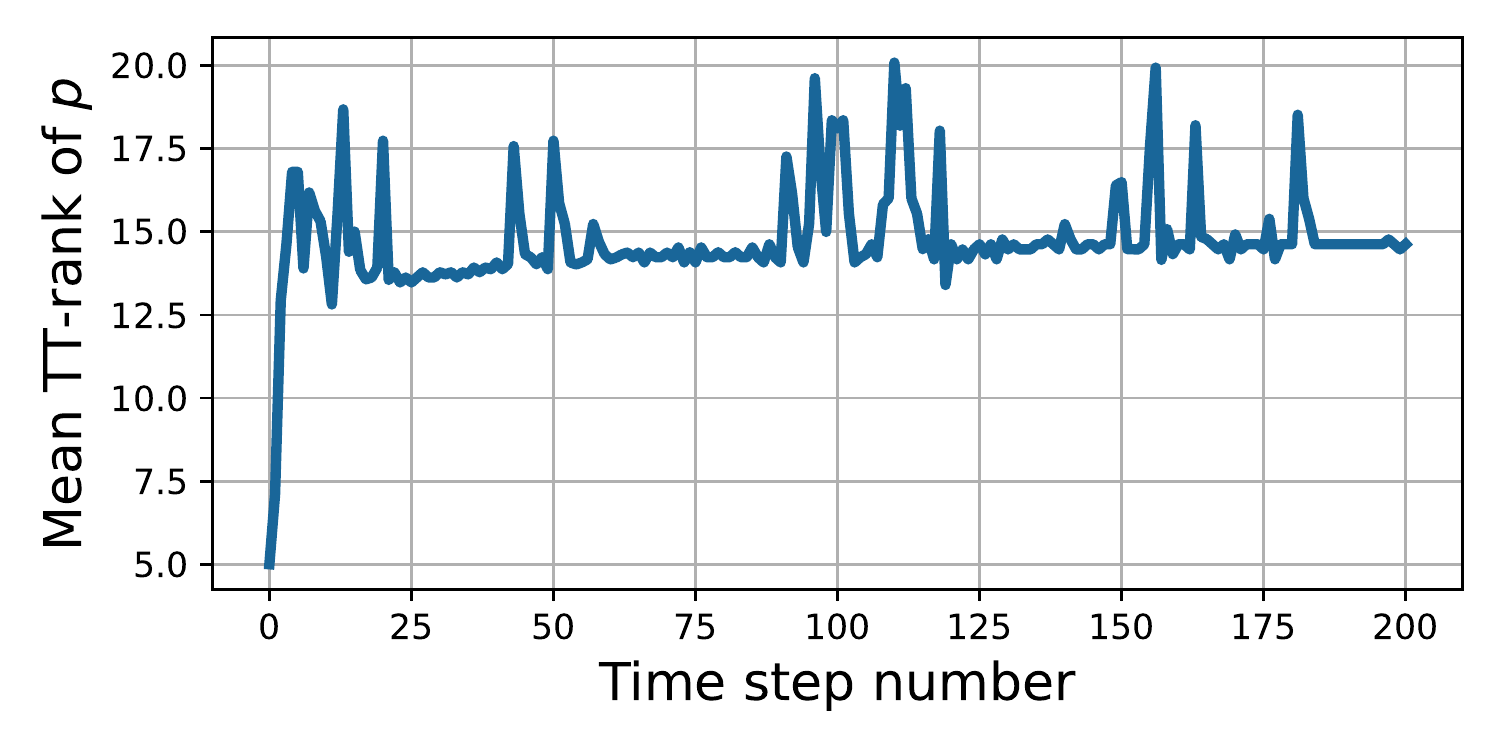}
    \caption{Dynamics of the mean TT-rank of the pressure vector during simulation of the T-mixer. 
    As the physical time increases, the flow behavior becomes more complex, which affects the increase in TT-rank.}
    \label{fig:rank}
\end{figure}

\section{Extension to a Quantum Computer}\label{quantum_extension}



This section outlines how utilizing a quantum computer can provide further benefits to this method. 
As we see, we can realize the whole numerical scheme as operations with Tensor Trains. 
However, if we observe the dynamics of TT-ranks during the iterations, we will see that they increase over time (Fig.~\ref{fig:rank}). 
This happens because the flow becomes more complex over time -- there are more correlations in the solution vector during the iterations. 
Thus, if we want to solve problems with more complex flows or geometric shapes, or solve the problem more precisely, then we will inevitably face an increase in TT-ranks and, accordingly, the complexity of the entire algorithm.

A quantum computer can help us resolve this problem.
Indeed, a quantum computer can do the following: multiply an initial state-vector, $X_0$, by matrices:
\[X = V_n V_{n - 1} ... V_2 V_1 X_0,\]
where $V_i$ are unitary matrices. 
Our assumption is that if we have a numerical scheme written as operations with Tensor Trains (as in our case), then we will be able to \textit{efficiently} create a $V_i$ matrix for each iteration step. 
By efficiency, we mean that a quantum circuit, which corresponds to each $V_i$ decomposed into single and two-qubit gates, is shallow and does not require full connectivity between the qubits. 
For example, any Tensor Train can be efficiently encoded into a quantum circuit \cite{MPS_preparation, two_qubits_MPS_encoding, QPrep, encoding}. 
We conjecture that other operations can also be extended, for example, multiplication by a TT-matrix seems to be the most native operation we are currently working on. 
The most important thing is that if we manage to implement all operations with tensor networks on a quantum computer, then we will be able to solve problems with any TT-ranks since the complexity of quantum computer algorithms will not depend on them at all.


\section{CONCLUSION}\label{conclusion}

To summarize, we propose to use the Tensor Train Finite Element Method for the solution of the Navier-Stokes equation in complex geometries. 
For this, we divide a complex domain into curvilinear quadrilaterals, transform them into quads, and generate local stiffness, mass and first-derivative matrices. 
Then, we assemble the global FEM matrices taking into account the connection of subdomains on the borders. 
We represent these matrices and initial vectors in the Quantized Tensor Train format in order to implement an explicit iterative scheme for the numerical solution of Navier-Stokes in the TT format. 
This allows us to exponentially reduce the memory consumption and gain an exponential speed up in comparison to the conventional Finite Element Method. 
In addition, the results show that the equations are solved correctly since the solutions coincide with the experimental data. 
Moreover, we can control the solution accuracy and lower it, if desired, in return for a higher speed.


We plan to develop an extended method
for curvilinear domains with analytic boundaries in 3D domains, wrapping it in a multi-purpose Finite Element toolbox. 
It will also possess a more accurate connection of sub-domains and faster reassembling of the convection term matrix $-(\vec{v} \cdot \nabla)$ on each iteration step.
This efficient and flexible toolbox will allow us to solve PDEs in more complex geometries much more efficiently, which can be applied to tasks in various fields requiring huge discretization that modern computers cannot handle.


In addition, we discuss how to extend this approach to a quantum computer. 
To do this, one needs to develop the methods that implement all operations with Tensor Trains on a quantum computer.
This will allow us to tackle more complex flows and achieve better accuracy since the complexity will no longer depend on the ranks of the solution.

\bibliography{lib}
\bibliographystyle{unsrt}
\end{document}